\documentclass[letter,12pt]{article}

\newcommand{\Z}{\mathcal{Z}}

\newcommand{\be}{\begin{equation}}
\newcommand{\ee}{\end{equation}}
\newcommand{\bea}{\begin{eqnarray}}
\newcommand{\eea}{\end{eqnarray}}
\newcommand{\ba}{\begin{eqnarray}}
\newcommand{\ea}{\end{eqnarray}}

\newcommand{\beq}{\begin{equation}}
\newcommand{\eeq}{\end{equation}}
\newcommand{\beqa}{\begin{eqnarray}}
\newcommand{\eeqa}{\end{eqnarray}}
\newcommand{\beqar}{\begin{eqnarray*}}
\newcommand{\eeqar}{\end{eqnarray*}}

\newcommand\al{\alpha}

\begin{document}
\title{Non-relativistic metrics from back-reacting fermions}
\author{Ling-Yan Hung$^{1}$\thanks{jhung@perimeterinstitute.ca}, Dileep P. Jatkar$^{1,2}$\thanks{djatkar@perimeterinstitute.ca} and Aninda Sinha$^{1}$\thanks{asinha@perimeterinstitute.ca} \\[0.4cm]
\it $ ^1$Perimeter Institute for Theoretical Physics\\
\it Waterloo, Ontario N2L 2Y5, Canada.\\[.5em]
 \it $ ^2$Harish-Chandra Research Institute\\
 \it Chhatnag Road, Jhusi,
 Allahabad 211019, India.
  }
\maketitle

\vspace{2cm}
\abstract{ It has recently been pointed out that under
  certain circumstances the back-reaction of charged, massive Dirac
  fermions causes important modifications to AdS$_2$ spacetimes
  arising as the near horizon geometry of extremal black holes. In a
  WKB approximation, the modified geometry becomes a non-relativistic
  Lifshitz spacetime. In three dimensions, it is known that
  integrating out charged, massive fermions gives rise to
  gravitational and Maxwell Chern-Simons terms. We show that
  Schr\"odinger (warped AdS$_3$) spacetimes exist as solutions to a
  gravitational and Maxwell Chern-Simons theory with a cosmological
  constant. Motivated by this, we look for warped AdS$_3$ or
  Schr\"odinger metrics as exact solutions to a fully back-reacted
  theory containing Dirac fermions in three and four dimensions. We
  work out the dynamical exponent in terms of the fermion mass and
  generalize this result to arbitrary dimensions.  }

\newpage

\tableofcontents
\section{Introduction} \label{intro} In recent times, there has been a
concerted effort to find real world applications of the AdS/CFT
correspondence. By extending the rules of the correspondence to finite
temperature, there are now well established prescriptions to calculate
retarded Green functions which enables us to extract transport
properties of interesting physical systems. Although none of these
systems, strictly speaking, describe the real world, it is hoped that
by studying these problems, one will gain intuition about real
physical systems which are for example in the same universality
class. On one hand, AdS/CFT methods have found applications to
particle physics, e.g., the viscosity of the strongly coupled quark
gluon plasma \cite{reviewsQGP}.  On the other,
these methods have been extended to studying table-top condensed
matter physics \cite{reviewscond}. It is fair to say that applications
of gauge/gravity duality will yield many more surprises.

One application that has garnered much attention lately is the
computation of fermion Green functions using AdS/CFT which yield
non-Fermi liquid type behaviour \cite{fermions}. In the simplest
example, one starts with a Reissner-Nordstr\"om AdS black hole and
considers the Dirac equation in this background. In the zero
temperature limit, the black hole becomes extremal and the near
horizon geometry contains an AdS$_2$ factor. This AdS$_2$ is thought
to play an important role in the existence of a Fermi surface. The
resulting fermion Green function shows non-Fermi liquid type behaviour
which is exciting in the context of connections with condensed matter
systems, e.g. high T$_c$ superconductivity. However, in the zero
temperature limit, the black hole and hence the corresponding dual
field theory, still has a large entropy which is far from what is
expected in the realistic situations.

Recently in \cite{hpst}, it has been pointed out, that in certain
regimes the back reaction of the fermions cannot be ignored. In fact,
the AdS$_2$ geometry has been argued to be modified into a ``Lifshitz"
\cite{Kachru:2008}
type and the associated entropy enigma is thought to disappear since
the horizon area is now zero. Motivated by this, we examine the
problem of back-reaction of Dirac fermions in 2+1 and 3+1 dimensional
gravitational theories in the context of AdS/CFT in some detail. The
2+1 dimensional bulk theory will describe 1+1 dimensional field
theories. In this context interesting non-Luttinger type behaviour has
been found in \cite{hs}. The 3+1 dimensional bulk theory will describe
interesting 2+1 dimensional field theories which are of relevance to
condensed matter physics related to high T$_c$ superconductivity. We
will restrict our attention to the simplest case where only the
metric, a $U(1)$ gauge field and a charged, massive Dirac fermion is
involved.

We begin with the discussion of a 2+1 dimensional bulk theory.  It is
known that \cite{bb,red,bpr} when one integrates out a charged fermion
in 2+1 dimensions, the effective action that one is led to contains
the Maxwell and gravitational Chern-Simons terms. Gravitational
Chern-Simons theories have been the subject of much attention recently
due to the connection with topologically massive gravity
(TMG)\cite{tmg}.  In TMG null warped AdS$_3$ have already been
found. The new finding in our case is their existence in the presence
of $U(1)$ Chern-Simons term $A\wedge F$.  We show that in addition to
the usual asymptotically AdS$_3$ spacetime, one also gets
Schr\"odinger spacetimes \cite{schr} whose dynamical exponent depends
on the Maxwell Chern-Simons or the gravitational Chern-Simons
terms. Similar solutions (called warped AdS$_3$) have already been
reported in \cite{nullads3} in the presence of the gravitational
Chern-Simons terms.  Inspired by this finding, we look for
Schr\"odinger spacetimes as exact solutions to back-reacted Dirac
fermions. Since fermions obey the Pauli exclusion principle, they
cannot be treated classically. We will treat the system of gravity,
gauge field and fermions in a semi-classical manner following
\cite{brillwheeler, ruffini}.  In this approach, the fermion stress
tensor and current appearing in the equations of motion are evaluated
as expectation values in some state $|Q\rangle$\footnote{This method
  is used quite routinely in nuclear physics.  See, {\it e.g.},
  \cite{Gambhir:1989mp}.}.  We will consider $|Q\rangle$ to be the
state made from $N$ fermions. It will turn out that the role of
$\hbar$ is played by $1/Nq$ so that for a fixed charge, taking the
large $N$ limit will correspond to making $\hbar$ small which is
needed for a semi-classical approximation to make sense. The solution
to the full problem in a self-consistent manner in general is a very
hard task. Typically one resorts to some approximation as in
\cite{ruffini} where a WKB approximation is used to derive the
Oppenheimer-Volkoff equations.  Quite remarkably, our problem will
turn out to be amenable to an exact solution! The key ingredient that
makes this possible is the fact that the fermion couples to the gauge
field. The equations of motion then lead to certain important
constraints which make an exact solution possible. In arbitrary bulk
dimensions, $D$, we will show that the dynamical exponent is given by
\be z=\pm 2 m L-(D-2)\,, \ee where $L$ is the length scale entering in
the definition of the cosmological constant in the usual way. Lifshitz
metrics are known to emerge as a WKB approximation \cite{hpst}.  It is
not known if they are \emph{exact} solutions to a system of
self-gravitating fermions as in the case of the Schr\"odinger
metrics. In the 2+1 dimensional case, as we will show, Lifshitz
metrics are not exact solutions to the equations of motion. However,
in higher dimensions we expect there to be more general ways of
solving the equations and it could be that other interesting solutions
exist. We leave this interesting question as an open
problem.\footnote{The fermionic stress tensor eq.(7.43) in \cite{hpst}
  can be shown not to lead to Schr\"odinger spacetimes.}

This paper is organized as follows. In section 2, we consider
Chern-Simons theories in 2+1 dimensions and show that Schr\"odinger
type metrics with specific dynamical exponents exist as solutions to
the system of equations. Motivated by this, we ask if similar metrics
exist as fully back-reacted solutions to Dirac fermions in section
3. We work out the dynamical exponent in arbitrary dimensions, i.e.,
where the original metric would have been AdS$_2 \times$ R$^{D-2}$.
We conclude with open questions in section 4.  In appendix A we spell
out our conventions and in Appendix B, we derive spin connections for
a certain class of metrics, which includes, AdS, Schr\"odinger and
Lifshitz metrics and set up the Dirac equation in these backgrounds.
In Appendix C we consider the stress tensor for a Dirac
fermion{\footnote{Dirac systems in AdS spacetimes have been studied
    also in \cite{bachelot}.}} in a background containing an AdS$_2$
factor.  We denote the charge of the fermion by $q$ and mass by $m$.
The AdS$_2$ arises as the near horizon geometry of some extremal
Reissner-Nordstr\"om black hole.  In particular, a gauge field $A_t$
needs to be turned on for Einstein's equations to be satisfied.  We
consider a charged, massive Dirac fermion in this background as a
probe and work out the stress tensor.  We find that quite generally an
off-diagonal component of the stress-tensor is turned on.  We show
that when $m<\sqrt{2 q^2+1}$, the back-reaction of fermions becomes
important. In appendix D, we study a quantum mechanical toy model in
path integral formalism which corresponds to coupling of macroscopic
fermi gas to a bosonic harmonic oscillator.  We integrate out the
fermions and obtain an effective action for the boson.  We show that
when there are a large number of fermions, the back reaction of fermions
is well approximated by replacing fermion number operators by their
expectation values evaluated in the $N$ fermion state.

\setcounter{equation}{0}
\section{Schr\"odinger spacetimes in Chern-Simons theories}
\label{sec:schr-spac-chern}

We will begin with the discussion of 2+1 dimensional
Einstein-Maxwell-Chern-Simons theory.  Our motivation for looking at
this theory came from the observation that for a probe fermion in
AdS$_2$ geometry, off-diagonal components of the stress tensor are
non-zero.  (This has been discussed in detail in appendix C.)  As a
result it may be expected that any non-relativistic modification of
the metric will be of the Schr\"odinger type rather than Lifshitz.  As
explained in the introduction, on integrating out fermions in 2+1
dimensions, one gets Maxwell and/or gravitational Chern-Simons
theories.  Rather than looking at the more complicated problem of
back-reaction of fermions, we wish to begin by investigating if
Schr\"odinger type solutions exist for an Maxwell and/or gravitational
Chern-Simons theories. Consider the action given in eq. (\ref{eq:19}).
The equations of motion for the metric read
\be
R_{\al\beta}-\frac{1}{2}g_{\al\beta}R-\frac{1}{L^2}g_{\al\beta}-
\frac{1}{2}(F_{\al \gamma}F_\beta^{\ \gamma}-\frac{1}{4}F^2
g_{\al\beta})+\mu_G C_{\al\beta}=0\,,
\ee
where $C^{\mu\nu}=\epsilon^{\mu \alpha \beta}\nabla_\alpha
(R^\nu_\beta-\frac{1}{4}\delta^\nu_\beta R)$ is the Cotton tensor.
The equations of motion for the gauge field read
\be
\nabla_\nu F^{\nu \mu}+2 \mu_F \epsilon^{\mu \nu\rho}F_{\nu\rho}=0\,.
\ee
It can be shown that the following Schr\"odinger\footnote{This
  nomenclature is motivated by the obvious similarity of the metric
  with the standard Schr\"odinger spacetimes which involve additional
  spatial directions. Strictly speaking this is an abuse of
  nomenclature since in our case there are no spatial directions which
  transform under dilatations. The conformal boundary of such
  spacetimes has been discussed in detail in \cite{confinf}. } (or
warped AdS$_3$ \cite{nullads3}) solutions exist
\begin{eqnarray}\label{csone}
ds^2&=&-r^z dt^2\mp 4 r dt dx+\frac{dr^2}{4r^2}\,,\\ \label{cstwo}
A_t &=& \frac{2}{z}[r^z z(z-1)(1\mp \mu_G\pm 2\mu_G
z)]^{1/2}\,,\\ \label{csthree}
z&=&0, 1, \mp 4\mu_F, \frac{\mu_G\mp1}{2\mu_G}\,,
\end{eqnarray}
To avoid clutter, we have set $L=1$. Here $z$ is the standard
dynamical exponent and $z=1$ corresponds to
the AdS solution, whereas $z=0$ corresponds to a chiral wave AdS solution.
The thing to note here is the existence of
Schr\"odinger solutions when either or both of the Chern Simons terms
exist. The existence of these solutions in the presence of the
gravitational Chern-Simons term can be anticipated from the general
discussion in \cite{amsv}. The case $z=2$, ($\mu_F=\mp 1/2$ or
$\mu_G=\mp 1/3$) is called the null-Warped AdS$_3$ \cite{nullads3}.

At this point it is worth noting that Chern-Simons terms are not
invariant under parity (P) and time reversal (T) transformations but are
invariant under combined transformation PT.  This symmetry is also
shared by the Schr\"odinger background, which under P or T
transformation swaps the sign of $dtdx$ term but the metric is
invariant under combined PT transformation.  However, in case of $z=1$
the metric is diffeomorphic to AdS metric and  preserves P and
T separately.  This does not contradict earlier conclusion because for
$z=1$, all components of gauge field vanish and in that case it is
natural to expect to recover P and T symmetry.

With a bit more work it can also be shown that a Lifshitz metric is
not a solution to the above set of equations. We begin with the ansatz
\begin{equation}
  \label{eq:3}
ds^2=-\frac{dt^2}{r^{2z}}+\frac{dr^2+dx^2}{r^2}\,, \quad A_t=\phi(r),
A_r=0, A_x=\chi(r)\,.
\end{equation}
Then the $tt$ component of the metric equations of motion leads to
$r^{2+2z}\phi'^2+r^4 \chi'^2=0$ while the $xx$ component leads to
$4-4z^2+r^{2+2z}\phi'^2+r^4 \chi'^2=0$.  Combining these two we are led
to $z=\pm 1$.  The solution $z=1$ is the usual AdS.  The choice $z=-1$
can be shown to lead to imaginary gauge fields. Thus we conclude
that Lifshitz is not a solution to this system.  We now turn to the
more complicated problem of considering back-reaction due to fermions.

\setcounter{equation}{0}
\section{Schr\"odinger spacetimes from fermions}
\label{sec:schr-spac-from}
If we compute the stress tensor for a probe fermion in an AdS$_2$
background as in appendix C, we find that it contains off-diagonal
components. This suggests that if their back-reaction is taken into
account an exact background geometry of the form of a Lifshitz
geometry, which is diagonal, is unlikely.  Given however, the
intuition from flat space that fermions and the Chern-Simons terms are
intimately related, and the similarity in structure of the
energy-momentum tensors of the gauge fields in the presence of the
gauge Chern-Simons term and that of the fermions, one is tempted to
conjecture that the back-reaction of charged fermions would also lead
to Schr\"odinger spacetimes, exactly as shown to happen, in the
previous section when the back-reaction of gauge fields in the
presence of Chern-Simons terms is taken into account.
To be explicit, the full Lagrangian we will be considering is (total
bulk spacetime dimensions is denoted by $D=d+1$)
\be
S= \frac{1}{2\ell_p^{d-1}} \int d^{d+1}x\sqrt{-g}\left[ R -2 \Lambda
-\frac{1}{4}F_{\mu\nu}F^{\mu\nu} + i(\bar\psi\Gamma^ae_a^\mu
  D_\mu\psi-m\bar\psi\psi)\right],
\ee
where, $\Lambda = -\frac{d(d-1)}{2L^2}$, the $\Gamma$-matrix
convention is given in Appendix B.
Note here that in general the fermion field will be a sum over modes
\begin{equation}
  \label{eq:15}
  \psi (r,t,x) = \sum_k  \left( a_{k} \psi_{a,k}(r,t,x)  + b^\dagger_k
\psi_{b,k}(r,t,x)\right) , \, \psi^\dagger (r,t,x) = \sum_k \left( a^\dagger_{k} \psi^
  \dagger_{a,k}(r,t,x)  + b_k  \psi_{b,k}^\dagger(r,t,x)\right)
\end{equation}
where $a_{k}, a_k^\dagger, b_k,b_k^\dagger$ denotes the anti-commuting
creation and annihilation operators of the corresponding fermion
and anti-fermion modes $\psi_{a/b,k}(r,t,x)$ respectively, $k$ denoting some
general quantum numbers of the modes, each mode $\psi_{a/b,k}(r,t,x)$ a
two-component spinor being a solution to the Dirac equation, and that
to lowest order ignoring quantum fluctuations of the background we
have,
\be\label{anticom}
\{a_{k_i}, a^\dagger_{k_j}\} = \hbar \delta_{k_i , k_j}= \{b_{k_i}, b^\dagger_{k_j}\},
\ee
while all other anti-commutations between operators vanish. Since the
background geometry itself is strongly back-reacted by the fermions,
in general it is slightly ambiguous to define what the vacuum would
be. We assume the existence of such a stable vacuum such that
\be
a_k|0\rangle = b_k|0\rangle =0.
\ee
Any general excited state would be of the form
\be
|Q \rangle_{general} = \prod_i a^\dagger_{k_i} \prod_j b^\dagger_{k_j}|0\rangle,
\ee
although they are not generally the $N$-fermion ground state at zero temperature, and
are thus not necessarily stable against decay.
The Einstein equation is, in
our conventions (Appendix A),
\be \label{Einstein}
R_{\al\beta}-\frac{1}{2}g_{\al\beta}R-\frac{d(d-1)}{2 L^2}g_{\al\beta}-
\frac{1}{2}(F_{\al \gamma}F_\beta^{\ \gamma}-\frac{1}{4}F^2
g_{\al\beta})= \frac{1}{2}\langle Q|T^f_{\al\beta}|Q\rangle\,,
\ee
where $T^f$ is as defined in (\ref{fermT}) and we reproduce here for convenience
\be
T^f_{\mu\nu} = \frac{1}{2}\left( - i \bar{\psi} \Gamma_{(\mu}D_{\nu)} \psi
+ i  \bar{\psi}\overleftarrow{D}_{(\mu} \Gamma_{\nu)}\psi \right).
\ee
This form of the stress tensor is given in \cite{birrell} and to the
best of our knowledge was first given in \cite{brillwheeler}. The covariant
derivative $D_\mu$ is defined in equation (C.2).
The Maxwell's equations are given by
\begin{equation}\label{Maxwells}
\nabla^\nu F_{\mu\nu}= \langle Q|j_\mu|Q\rangle\, \qquad j_\mu
= -q \bar{\psi}\Gamma_\mu\psi.
\end{equation}

Note that in the above expressions, the fermion stress tensor and
electric currents that back-react on the background are really the
expectation values of the corresponding operators, evaluated on the state
$|Q\rangle$. At zero temperature, the ground state of $N$ fermions
would correspond to piling up the various modes, starting from the
bottom mode. Naively, both the currents and the fermion stress tensor
would take roughly the form $\sum_k a^{\dagger}_{k}
a_{k}|\psi_k(r,t,x)|^2 \, $, whose expectation value in the state
$|Q\rangle$, would involve a sum over contributions of modes that are
occupied in $|Q\rangle$.  However, here we should note that in general one is confronted with a diverging expectation value due to
contributions even of unfilled modes. It is ambiguous to define
normal-ordered operators in curved space, since vacuum energy
gravitates as well \cite{birrell}.  Here we approach the problem via analytic
continuation \cite{birrell}. As we will see in the following section, the modes that
are consistent with the background symmetry and the Gauss's law
constraints are simply chiral plane waves along $t$
in Schr\"odinger space, with trivial radial dependence. Therefore they
contribute in the expectation value of the current or stress tensor via $\langle Q|
\sum_i a_i^\dagger a_i|Q\rangle$ simply as $\sum_i^N(1)=N$ and thus
precisely in the same way the unoccupied anti-fermion modes, which we
have infinitely many of them will contribute via
\be\label{regulate}
\langle Q| \sum_i b_ib_i^\dagger|Q\rangle = 1+1+1 ... = \zeta(0) = -\frac{1}{2},
\ee
where in the second equality we attempt to regularize it via analytic
continuation.  Adopting this approach would correspond to shifting $N$
by $-\frac{1}{2}$. Note that this is sub-leading in $N$, and in the
large $N$ limit, which is the main focus of this paper, we will ignore
this term.

The analysis that we present below is very similar in spirit to that
in \cite{ruffini} for a system of self-gravitating fermions.  The
essential  steps in this are as follows:
\begin{enumerate}
\item We want to find a particular set of fields involving the metric,
  gauge field and fermions such that it is an exact solution to the equations of
  motion given above.
\item In \cite{ruffini}, a certain ansatz for the metric compatible
  with some pre-supposed symmetries is taken. The Dirac equation is
  solved in this background for a complete set of modes.
\item With these solutions, one works out the stress tensor $T^f$
  entering Einstein's equations, taking care of the antisymmetric
  nature of the fermions.
\item The full set of equations is now solved self consistently.
\end{enumerate}
In the regime of large quantum numbers and using a WKB approximation
\cite{ruffini} recovers the Oppenheimer-Volkoff equations\footnote{In
  \cite{holoneu}, a similar analysis was used to study holographic
  neutron stars.}. In \cite{hpst}, a WKB approximation was used to
obtain a Lifshitz solution. We will not make any such approximations
in what we do. In addition to the metric and fermions as in
\cite{ruffini} we also have a gauge field which makes the calculation
somewhat more complicated.  To keep life simple, we will begin with
the ansatz for a Schr\"odinger spacetime with an undetermined
dynamical exponent. Then we will ask if there exist self-consistent
set of solutions. To be compatible with the symmetries of the problem,
we will assume that the gauge field is only dependent on the radial
coordinate. At this stage, we note that
the stress tensor for fermions has an $\hbar$ \cite{ruffini}. We will
find that $1/N q$ plays the role of $\hbar$ in what we do.  A quick
way to see that this is true is that in our solution, the stress
tensor from fermions is of the same order as that of the Maxwell
fields and the Einstein tensor, not surprisingly since we found a
fully back-reacted solution. However, the general form of the charged
fermion stress tensor is roughly $T^f \sim \hbar N q$, using the
anti-commutation relation of the operators. Therefore our solution
naturally forces $N q \sim 1/\hbar$.
It is interesting to note here that for small $N q$, it would imply
that $\hbar$ is large and that a classical approximation would not
stand. This conforms with our intuition that a classical approximation
improves when the fermi gas becomes macroscopic. To find further support for
our approach, we present a path-integral calculation of a toy-model consisting of
fermions coupled to bosons in Appendix D. We found that quite generically in the limit
of large fermion number $N$, our semi-classical treatment adopted here is a
good approximation.

\subsection{2+1 Dimensions}
\label{sec:2+1-dimensions}

Like the Chern-Simons terms, massive fermions in 2+1 dimensions also
break P and T symmetry.  We therefore expect the result here to be
similar to what we got in the presence of Chern-Simons terms.
Anticipating that the back-reacted exact geometry is going to be given
by the Schr\"odinger metric, we start with the ansatz metric exactly
as in the previous section, namely
\begin{equation}
\label{sch3d}
ds^2=L^2(-r^z dt^2-4 \epsilon r dt dx+\frac{dr^2}{4r^2}),
\end{equation}
where $\epsilon = \pm1$, and leave the gauge fields and the index $z$
to be determined by the equations of motion. The three sets of
equations involved are the Maxwell equations, the Einstein equations
and the Dirac equations, to be solved simultaneously. We will, unless
otherwise specified, set $L=1$ to avoid clutter in the equations.

To begin with, we compute the spin-connection necessary to build the
Dirac equations.
Choosing explicitly the vierbein as
\begin{equation}
e^0=r^{-\frac{z}{2}} (r^z dt + 2 \epsilon r dx), \qquad e^1 = 2
\epsilon r^{1-\frac{z}{2}} dx, \qquad e^d = \frac{1}{2r}dr,
\end{equation}
the corresponding spin-connection is
\be \label{schrospin}
\omega^{01} = \frac{z-1}{2} dr, \qquad \omega^{0d}=
r^{-\frac{z}{2}}(r^z z dt + 2\epsilon r dx),\qquad \omega^{1d}
=r^{-\frac{z}{2}}(r^z (z-1) dt + 2\epsilon r dx).
\ee
Considering a particular mode of the Dirac spinor,
\be
\psi_{\omega,k}(r,t,x) = e^{-i \omega t -i k x}\left(\begin{array}{c}
\psi_{+,\omega,k} (r)\\ \psi_{-,\omega,k}(r)
\end{array}\right),
\ee
the Dirac equation for a fermion of mass $m$ takes the form
\begin{eqnarray}
  \label{eq:14}
&&  -(2 + 4 i q r A_r(r)) \psi_{-,\omega,k}(r) + 2 i r^{-z/2}
( \omega - q A_t(r)) \{\psi_{-,\omega,k}(r) - \psi_{+,\omega,k}(r)\} \nonumber \\
&&+ [1 + 2 m - z +
 i r^{(-1 + z/2)} (k - q A_x(r))] \psi_{+,\omega,k}(r) - 4 r \psi'_{-,\omega,k} (r)=0 ,\\
&&- 2 i r^{-z/2} (\omega - q A_t(r)) \{\psi_{-,\omega,k}(r) - \psi_{+,\omega,k}(r)\} +
 (2 + 4 i q r A_r(r)) \psi_{+,\omega,k}(r) \nonumber \\
&& -[1+ 2 m - z -i r^{(-1 + z/2)} (k - q A_x(r))]
\psi_{-,\omega,k}(r) + 4r \psi'_{+,\omega,k} (r)= 0.
\end{eqnarray}

It is simplest to start examining the Maxwell's equations (\ref{Maxwells}).
We shall begin by picking the gauge
\be
A_r =0.
\ee
We will be looking at radially symmetric solution such that the metric and the
gauge fields contain only $r$ dependence. Given our gauge choice, the Maxwell's equation
for the $A_r$ component becomes purely a constraint on the vanishing
of the $r$-component of the
fermion current. That gives, explicitly,
\be \label{jr}
j_r = -q \epsilon \frac{(\hat{\psi}^\dagger_- \hat{\psi}_- -
\hat{\psi}_+^\dagger \hat{\psi}_+)}{2r},
\ee
where we have defined the operators $\hat{\psi}_\pm$, to be
distinguished from the modes $\psi_{\pm,\omega,k}$
\be
\psi = \left(\begin{array}{c}\hat{\psi}_+ \\\hat{\psi}_-
\end{array}\right),\qquad \hat{\psi}_\pm  = \sum_i a_i
e^{-i( \omega_i t-k x)} \psi^a_{\pm,\omega,k} + b^\dag_i
e^{i( \omega_i t-k x)} \psi^b_{\pm,\omega,k}.
\ee
At the same time, to preserve the symmetry of the system, we would
like to keep the magnetic field zero i.e. $F_{xr}=0$,
we allow only for
\be\label{solstep2}
A_x(r) = a^x_1,
\ee
for some constant $a^x_1$, which, from the gauge equations, further give
\begin{equation}
  \label{eq:28}
  A_x'(r) + r A_x''(r) = - q \epsilon ((r^{-z/2})
\langle Q|(\hat{\psi}^\dagger_- - \hat{\psi}^\dagger_+) (\hat{\psi}_-
- \hat{\psi}_+))|Q\rangle = 0.
\end{equation}
To satisfy both requirements (\ref{jr},\ref{eq:28}), in general we have
\be\label{solstep1}
\psi^{a}_{-,\omega,k}(r) = \psi^{a}_{+,\omega,k}(r) , \qquad \psi^{b}_{-,\omega,k}(r) = \psi^{b}_{+,\omega,k}(r),
\ee
for each contributing mode occupied in state vector $|Q\rangle$.
%

Now going back to the Dirac equations, given the relation (\ref{solstep1}) the Dirac equations reduce to
\be
 r ((1 - 2  m + z) \psi^{a,b}_{+,\omega,k} + 4 r{\psi^{a,b}_{+,\omega,k}}')
\pm i r^{\frac{z}{2}} (k - q A_x) \psi^{a,b}_{+,\omega,k}=0.
\ee

It is important to note that with the relation (\ref{solstep1}), the
$A_t$ and time-derivatives have completely dropped out from the Dirac
equation. This implies that all the modes with different $\omega$
share the same radial wave-function, which also means
\be
\psi^a_{\pm,\omega,k}=\psi^b_{\pm,\omega,k}.
\ee
Also, the term proportional to $(k - q A_x)$ renders the two equations
inconsistent. To obtain non-trivial solutions, it implies
\be\label{Ax}
a^x_1 = \frac{k}{q},
\ee
which cannot be satisfied generally for an arbitrary $k$-mode in a
given background of some $a^x_1$. This implies that the modes consistent with
the background are those with arbitrary $\omega$ but fixed $k$ such
that (\ref{Ax}) is satisfied. These solutions however are gauge
equivalent to having $k=A_x=0$, via the gauge transformation $\chi =
-\frac{k}{q} x$ i.e.
$
\psi \to e^{-i k x} \psi,
$
such that correspondingly
$
A_x \to A_x -\frac{k}{q}.
$
We will thus from now on set
$
A_x = k =0.
$
i.e., in full generality, we have
\be\label{modeexpand}
\psi(t,r) =
 \left(\begin{array}{c} 1\\1\end{array}\right) \sum_i
(a_{\omega_i} e^{-i\omega_i t}\Psi_{+}(r) + b^\dagger_{\omega_i} e^{i\omega_i t}\Psi_{+}(r) ),
\ee
where we have
\be
\psi^a_{\pm,\omega,0}=\psi^b_{\pm,\omega,0}=\Psi_{+}(r).
\ee

To avoid complications, for the moment we
treat the spectrum as discrete. This does not alter our results in any
significant way even when the continuous limit is taken.
We would like to consider the back-reaction of a large number of fermions. This
can be done by building explicitly a fermi gas by constructing the
$N$-fermion state vector $|Q\rangle$
\be
|Q\rangle = \prod_i^N a^\dagger_{\omega_i}|0\rangle.
\ee
Let us emphasize here that $N$ is the number of modes excited, which is \emph{infinity}
in a continuum limit for finite fermi-energy. The fermion density distribution $\rho$
however is smooth.
As we have demonstrated above, since modes of different $\omega$ have
the same $r$-wave function, evaluating the expectation value of the
stress tensor and the current on  $|Q\rangle$, would be proportional
to the term, schematically given by
\be
\langle Q|\hat{ \psi}_+^\dagger \hat{\psi}_+ |Q\rangle =
|\Psi_{+}(r)|^2  \langle Q| \sum_i^N a^\dagger_{\omega_i}a_{\omega_i}
+ \sum_i^\infty b^\dagger_{\omega_i}b_{\omega_i}  |Q\rangle =
\hbar (N-\frac{1}{2}) |\Psi_{+}(r)|^2,
\ee
where the unfilled-anti-fermionic states contribution $b b^\dag$ has
been regularized as discussed in (\ref{regulate}). Since they are
sub-leading in $N$ after regularization, we will not include them
explicitly in the rest of our discussion.  Note that in the
expectation values of the operators, cross terms occurring in the
product $\hat{\psi}^\dagger \hat{\psi}$ vanishes,  and as a result
both the stress tensor and the currents have no $x,t$ dependence.
Returning to the Dirac equation, one can then readily solve it to give
\be
 \Psi_{+}(r) = p_1 r^{\delta_p},\qquad \delta_p = \frac{1}{4} (-1 + 2 m - z),
\ee
for some suitable overall constant normalization $p_1$\footnote{Note that because of the
way we defined the measure of the frequency sum in (\ref{modeexpand}), the normalization $p_1$ should
be $\omega$ independent to be consistent with the anti-commutation relations (\ref{anticom}) and subsequently
canonical anti-commutation relation between the $\psi$ and $\bar{\psi}$.}.
The Maxwell equations is left with one component determining $A_t$.
The expectation of the $t$-component of the current is given by
\be
\langle Q| j_t|Q\rangle = -q r^{\frac{z}{2}}\langle Q| (|\hat{\psi}_-
|^2+ |\hat{\psi}_+|^2) |Q\rangle = -2r^{\frac{z}{2}+ 2\delta_p}Nq\hbar|p_1|^2.
\ee
Substituting in the Maxwell equations we have
\be
\hbar N|p_1|^2 q r^{\frac{1}{2}(z+ 4 \delta_p)} - 2 r A_t' - 2r^2 A_t''=0,
\ee
which gives,
\be\label{At}
A_t= \hbar \frac{2 N|p_1|^2 q r^{\frac{1}{2} (z + 4 \delta_p)}}{(z +
4 \delta_p)^2} + c^t_2 +  c^t_1 \log r,
\ee
for some constants $c^t_i$. The log term would appear in the energy-momentum tensor as
a lone term. We therefore  set
\be
c^t_1 =0.
\ee
Returning to the Einstein equations, we inspect specifically the $tt$ component.
Evaluating the LHS of eq. (\ref{Einstein}), we get
\be
R_{tt}-\frac{1}{2}g_{tt}R-g_{tt}-
\frac{1}{2}(F_{t\gamma}F_t^{\ \gamma}-\frac{1}{4}F^2
g_{tt}) = 2(r^z (-1 + z) z - r^2 A_t'^2),
\ee
and the expectation value of the $tt$-component of the fermionic
energy-momentum tensor evaluated on the state $|Q\rangle$,
\begin{eqnarray}
\langle Q| T^f_{tt}|Q\rangle &=& -2  r^{\frac{z}{2}} (
\langle Q | \hat{\psi}_+\partial_t \hat{\psi}_+|Q
\rangle- q A_t \langle Q | \hat{\psi}_+ \hat{\psi}_+|Q \rangle) \nonumber \\
&=& -2  r^{\frac{z}{2}+2\delta_p}\hbar (\sum_i^N |p_1|^2
\omega_i - q A_t\sum_i^N |p_1|^2) \nonumber \\
&=& -2r^{\frac{z}{2}+2\delta_p}|p_1|^2\hbar(\sum_i^N \omega_i  - q N ( c^t_2-
\hbar \frac{2 N|p_1|^2 q r^{\frac{1}{2} (z + 4 \delta_p)}}{(z +
4 \delta_p)^2})),
\end{eqnarray}
which, together with (\ref{At}), are substituted in to the Einstein
equation (\ref{Einstein}). The $tt$-component of the equation is thus
left with three different terms.  The
metric contributes a term that goes like $r^z$, and the gauge and
fermion stress tensor contain two different terms:
one contributes to a $r^{z+ 4\delta_p}$, and another term proportional
to $q N c^t_2 - \sum_i^N \omega_i$
goes like $\sim r^{\frac{z}{2} + 2\delta_p}$. The only way to obtain a
non-trivial solution is to switch this latter term off.
Therefore we have,
\be\label{chemicalpot}
c^t_2 = \frac{\sum_i^N \omega_i}{qN}= \sum_{n=0}^{N-1} \frac{n\pi}{N q l} = \frac{(N-1)N\pi}
{2 q N l} \sim \frac{\omega_F}{2 q},
\ee
where $l$ is the regulated size of the time direction.
To obtain non-vanishing spinor solution, one is forced to take
$\delta_p =0$. This relates the index $z$ to the fermion mass,
\be\label{zzsol}
z = 2 m L-1,
\ee
where we have restored $L$.
With that the Einstein equation is thus reduced simply to an algebraic
equation constraining $p_1$
\be \label{Einsteintt}
-2 N^2 \hbar^2 |p_1|^4 q^2 + (z-1) z^3 =0,
\ee
which readily yields
\be\label{normsol}
|p_1|^2 =  \frac{(( m L-1) (2 m L-1)^3)^{\frac{1}{2}}}{\hbar N q},
\ee
where we have restored the radius $L$ here. However, $p_1$ is the normalization of
our plane wave solutions. It has to take a fixed value. The normalization of the wave
function is determined by
\be
\int d^3 x  \Psi_+^\dag(r) \Psi_+(r)= |p_1|^2 V = 1,
\ee
where $V$ is the regulated volume.

Therefore, the relation (\ref{normsol}) fixes the value of $N$ that would back-react to give rise to the particular geometry and electric fields.i.e.
\be
 \frac{(( m L-1) (2 m L-1)^3)^{\frac{1}{2}}}{\hbar N q} = \frac{1}{V},
\ee
which gives
\be
\frac{N q \hbar}{V} = \rho \hbar = (( m L-1) (2 m L-1)^3)^{\frac{1}{2}} \sim \mathcal{O}(1),
\ee
where $\rho$ is the charge density. Einstein equation dictates that $\hbar \rho \sim 1$ which is expected from our intuitive argument presented at the beginning of the section.

Being a constant solution, the $tr$-component of the fermion energy momentum tensor
\be
T^f_{tr} = \frac{i r^z}{2}( \hat{\psi}^{\dagger'}_+ \hat{\psi}_+ - \hat{\psi}^{\dagger}_+ \hat{\psi}_+ '),
\ee
which is a lone term in the Einstein equation also vanishes. Similarly
replacing $m\rightarrow -m$ we will get $z=-2 m L-1$.

There are a few interesting limits to take. One could for example consider
taking the limit $m L\to \frac{1}{2}$ such that $z\to 0$. Staring at (\ref{Einsteintt},\ref{normsol}) it
means a non-trivial fermion solution is only recovered for $q\to0$ at the same rate. In this limit however
the gauge component $A_t$ becomes a constant, but in fact an infinite constant since $c^t_2$
is proportional to $1/q$. Remarkably this is precisely what happens in the gauge-gravitational Chern-Simons
analysis when we take the limit $z\to 0$ and $\mu_F \to 0$ in equation (\ref{csthree}).
Another possibility to consider is that for $q=0$ to start with, then the
homogenous solution in the Maxwell's equations, namely the log term, could have been
allowed without entering into Einstein equation through the fermion stress tensor. Indeed
we checked that this is a solution, provided that $m=\frac{1}{2}$ and $z=0,1$.
When $z=1$ the fermion radial wave-function is in fact not a constant but goes like
\be
\Psi_+(r) \sim r^{-1}.
\ee
When one compares with the gauge-gravitational Chern-Simons setting, where one analogously takes
$\mu_F=0$ and allows for the homogenous solution $A_t\sim c^t_2 + c^t_1 \log r $, it turns out
$z=0,1$ are again solutions, provided that one has to take
\be
\mu_G (z-1)z \sim {c^t_1}^2
\ee
in either case. i.e. $\mu_G$ curiously approaches infinity. There appears to be a one-to-one correspondence
between the gauge-gravitational Chern-Simons theory and the Einstein-Maxwell-Dirac system, as would
be expected if Chern-Simons terms should be recovered by integrating out fermions.
Although our results suggest that the relationship between the value of the Chern-Simons couplings
and the fermion mass and charge depart from that in flat space.


\subsection{3+1 Dimensions and beyond}
\label{sec:3+1-dimensions}
The procedure above can be simply generalized to higher
dimensions. Consider for concreteness the case $D=3+1$.
In this case the metric ansatz we use is the 4-d Schr\"odinger metric,
\begin{equation}
ds^2=L^2(-r^z dt^2-4 \epsilon r dt dx + r dy^2 +\frac{dr^2}{4r^2}),
\end{equation}
where we will again set $L=1$ from now on.
Using the same choice of vierbeins as in $2+1$ with the addition of
\be
e^{3} = \sqrt{r} dy ,
\ee
where the hat denotes tangent coordinates. The spin-connections will
be given again by (\ref{schrospin}), with only the new addition of
\be\label{schrospin2}
\omega^{d3} = -\sqrt{r} dy.
\ee
Choosing again the Dirac spinor to be
\be
\psi = \sum_{i,\chi} e^{-i (\omega t - k_1 x - k_2 y)}a^\chi_{\omega,k_1,k_2} \psi^\chi_{\omega,k_1,k_2} + e^{i (\omega t - k_1 x - k_2 y)} {{b^{\chi}}^{\dag}_{\omega,k_1,k_2}}\psi^\chi_{\omega,k_1,k_2},
\ee
where $\chi =\{1,2,3,4\}$ are spinor polarization indices,
\be
\psi^\chi_{\omega,k_1,k_2} = \left(\begin{array}{c}
\psi^\chi_{+,\omega,k_1,k_2}
    (r)\\  \psi^\chi_{-,\omega,k_1,k_2} (r) \end{array}\right) ,\qquad
\psi^\chi_{\pm,\omega,k_1,k_2} = \left(\begin{array}{c}
\psi^{\chi,1}_{\pm,\omega,k_1,k_2} \\ \psi^{\chi,2}_{\pm,
\omega,k_1,k_2},
\end{array}\right),
\ee
and we take as before $k_i=0, \, i=1,2$. For clear notations, we will
omit the polarization index $\chi$ in the following,
only to look explicitly for the non-trivial spinor polarization that
would solve the system of equations.
Assuming only $r$ dependence and going through similar analysis as in
the previous section leads to very simple results. Starting again with
the Maxwell equations to obtain constraints on the spinors, the
equation for $A_r$ gives
\be
\hat{\psi^2}^\dag_+ \hat{\psi^1}_{-}+\hat{\psi^1}^\dag_+
\hat{\psi^2}_{-} - \hat{\psi^1}^\dag_- \hat{\psi^2}_{+}-
\hat{\psi^1}^\dag_- \hat{\psi^2}_{+} =0,
\ee
where as in the previous section we are again denoting the components
of the $\psi$ field operator with a hat, to be distinguished from the
modes.
The simplest solution is again taking, for each mode
\be
\psi^2_{+,\omega,k_1,k_2} = \psi^2_{-,\omega,k_1,k_2}=0.
\ee
This choice also easily ensures the vanishing of the source
currents for $A_x$ and $A_y$.
Subsequently the Einstein equation and the Maxwell equations would
again imply
\be
A_r = A_x=A_y=0.
\ee
Fermions taking constant value is forced to be a solution
\be
\psi^1_{\pm,\omega,k_1,k_2} = p^1_\pm,
\ee
for some normalization constants $p^1_\pm$,
and we arrive at
\be
A_t = c^t_1 \log r + c^t_2 + \langle Q|\frac{q (
\hat{\psi}^{1\dagger }_- \hat{\psi}^{1}_- + \hat{\psi}^{1\dagger }_+
\hat{\psi}^{1}_+) r^{\frac{z}{2}}}{z (1 + z)}|Q\rangle ,
\ee
where
\be
|Q\rangle = \prod_i^{N_+} a^\dag_{+,\omega_i} \prod_i^{N_-}
a^\dag_{-,\omega_i}|0\rangle,\qquad \langle Q |
\hat{\psi}^{1\dagger }_\pm \hat{\psi}^{1}_\pm |Q \rangle
= N_\pm |p^1_\pm|^2,
\ee
and $a^\dag_{\pm,\omega_i}$ denotes the creation operators of the
fermion modes $\psi^1_{\pm,\omega_i}$ respectively.
$N_\pm$ are the fermion numbers for the $\pm$ modes respectively.
As in $d=2+1$, Einstein equation dictates that $c^t_1=0$ and
\be
(N_+ |p^1_+|^2  + N_-|p^1_-|^2) c^t_2 = \frac{1}{q}
\left(\sum_{i} \omega_{+,i} |p^1_+|^2 + \sum_{j}
\omega_{-,j} |p^1_-|^2 \right).
\ee
The two remaining Dirac equations for constant fermions cannot
be satisfied simultaneously for arbitrary massive fermions when both
$\psi^1_+$ and $\psi^1_-$ are non-vanishing.
Explicitly, the remaining Dirac equations are given by
\be
p^1_- (2 + 2 m + z)=0, \qquad p^1_+ (2 - 2 m + z)=0.
\ee
We are thus left with three different possibilities.
\subsubsection{Massless fermions }
For massless fermions,
\be
z=-2,
\ee
and the remaining Einstein equation is
\be
\frac{3 (12 - \langle Q|(\hat\psi^{1\dagger }_- \hat\psi^{1}_- +
\hat\psi^{1\dagger }_+
\hat\psi^{1}_+)^2 q^2)|Q\rangle}{2 r^2}=0,
\ee
which relates $N_-$ to $N_+$.
\subsubsection{Massive fermions }
For massive fermions, there are two independent sets of solutions.
\be
p^1_+ =0, \qquad N_-|p^{1}_-|^2 = \pm \frac{2 \sqrt{(9 + 18 m
+ 8 m^2)(m+1) (2 m+1)^2}}{q\sqrt{4m+3}},
\ee
and
\be
z = 2mL -2.
\ee
This solution applies when $m\ne -\frac{3}{4}$. When $m$ actually
takes that value, $\psi^1_-$ becomes unconstrained because the
Einstein equation vanishes without further constraint on $\psi^1_-$.
Similarly one could have
\be
p^1_-=0,\qquad N_+|p^{1}_+|^2 = \pm \frac{2 \sqrt{(9 - 18 m
+ 8 m^2)( m-1)( 2 m-1)^2}}{q\sqrt{4 m-3 }}
\ee
where
\be
z = -2mL -2
\ee
and $m \ne \frac{3}{4}$. Otherwise $N_+$ is unconstrained.

Here we have an interesting pattern to note. When $D=2+1$, the
resultant index $z = 2mL -1$ and at $D=3+1$ that gives, depending on
chirality, $\pm 2 m L -2$. It turns out that exact solutions of
charged constant spinor solutions and Schr\"odinger metrics can be
found in arbitrary dimensions.  In $D=4+1$ for example we have
\be
z = \pm2 m L -3.
\ee
There is a clear pattern in the constraint on $z$ as we go to higher
dimension, namely
\be
z = \pm2mL - (D-2).
\ee
The reason it gets shifted in this particular manner is that these
constraints follow from requiring the coefficient multiplying $\psi$
in the Dirac equation to vanish, for the existence of constant spinor
solutions. The mass term enters in the same way for arbitrary
dimensions, but for each additional dimension we get an extra
spin-connection component, as illustrated in (\ref{schrospin2}), which
enters into the Dirac equation, after multiplying by the curved space
gamma matrices, simply as a constant. Therefore as we move onto higher
dimensions the constraint on $z$ get recurrently shifted by $-1$,
starting from the value when $d=2+1$
\footnote{Note also that in all these solutions
$
T^t_{\ t} - T^r_{\ r} =0.
$}.
The same chiral modes that are obtained in the previous section appear
also here. However, in dimensions higher than $d=2+1$, spinor
dimensions increase such that the constraints from Maxwell's equations
become a much weaker restriction and it should then be possible
generally to switch on momentum in other directions. As a
result our solution would not be able to capture the back-reaction of a
general state where modes with non-trivial momenta $k$ are occupied.

\subsection{Can there be an exact Lifshitz solution?}
Given our analysis with the Chern-Simons theory and the form of the
probe fermion stress tensor, it would be interesting to look for {\it
  exact} Lifshitz solutions.  However, in the $k_\mu=0$ case in 2+1
dimensions we can show that this is not possible. Note that in light
of our discussion of the semi-classical analysis, this will only make
sense if the charge $q$ is large compared to the mass. The following should only be treated
schematically and we leave a more rigorous analysis for future work.

We will again start with the Einstein-Maxwell-Dirac system in 2+1
dimensions and look for an exact solution to the equations of motion
in the background given in eq.(\ref{eq:3}).  The set of equations are
\begin{eqnarray}
  \label{eq:4}
  \Gamma^\mu D_\mu \psi -m \psi =0, \qquad
   \nabla^\nu F_{\mu\nu} = j_\mu, &&\\
   R_{\alpha\beta} - \frac{1}{2}g_{\alpha\beta} R -
  \frac{d(d-1)}{2L^2} g_{\alpha\beta} -
  \frac{1}{2}(F_{\alpha\gamma}F_{\beta}^{\ \gamma} - \frac{1}{4}F^2
  g_{\alpha\beta}) = \frac{1}{2} T^f_{\alpha\beta}.&&
\end{eqnarray}
Since we are looking for field configurations depending only on $r$,
the left hand side of the $A_r$ equation of motion vanishes
trivially.  This imposes a chirality constraint on the components of
the Dirac fermion, namely,
\begin{equation}
  \label{eq:5}
  \psi_+ = \psi_-\, .
\end{equation}
Inserting this condition back into the Dirac equation, we get two
equations for $\psi_+$, which are inconsistent unless
\begin{equation}
  \label{eq:6}
  A_x = - r^{z-1} A_t\, .
\end{equation}
Using this condition the Dirac equation can be solved,
\begin{equation}
  \label{eq:12}
  \psi_+(r) = \sqrt{r} p_1 r^{\frac{z-2m}{2}}\, .
\end{equation}
We can then substitute this solution and the relation between $A_x$
and $A_t$, in the Einstein equation.  The $x$-$r$ component of the
Einstein equation then implies that $\psi_+$ is a real function and as
a result the integration constant $p_1$ is real.  Having done this, we
can solve the $A_t$ component of the Maxwell equation, which yields
\begin{equation}
  \label{eq:13}
  A_t(r) = c_2 + \frac{c_1 r^{1-z}}{1-z}  - \frac{2 p_1^2 q
    r^{1-2m}}{(1-2m)(2m-z)}\, .
\end{equation}
Finally we can solve for the remaining components of the Einstein
equations.  Although we have four equations, only two of them are
independent.  After having already determined fermion and gauge
fields, these two equations are purely algebraic in nature and they
constrain the remaining undetermined integration constants, $c_1$,
$c_2$ and $p_1$ and the parameter $z$.  Unfortunately, there is no
consistent choice of these constants that can satisfy both equations.
The same analysis can be applied generally for $k_\mu \ne 0$, which
suggests that Lifshitz spacetimes are not exact solution to the
equations of motion.  This however does not contradict the fact that
in the WKB limit it becomes an approximate solution.

\setcounter{equation}{0}
\section{Discussion}
\label{sec:discussion}

Motivated by results of \cite{hpst}, we studied a system of
self-gravitating charged Dirac fermions and showed that
non-relativistic Schr\"odinger metrics exist as exact solutions \footnote{It is important to note that the fermion gas we are considering
is a smooth density distribution, which alleviates Coulomb repulsion particularly
along the $x$-direction, that otherwise would have made the gas unstable.}.  In
2+1 dimensions, this was anticipated by studying the three dimensional
Einstein-Maxwell-Chern-Simons system.  Although the gauge Chern-Simons
term does not couple to the metric, it affects the background by
modifying the Maxwell equations.  We found that in the presence of
gauge and/or gravitational Chern-Simons term, one generically finds
Schr\"odinger spacetime as a suitable background.

One may be worried that by treating the problem involving fermions semi-classically, we could be missing important higher order effects. For starters, the effects of quantum gravity are under control as we are implicitly assuming that the gravitational coupling is small. Next, one could be worried that since there are a large number of fermions, there could be an enhancement of higher loop effects by their presence. However, as demonstrated in a general toy-model in Appendix D,
as long as the number of modes $N$ excited is large, the \emph{tree} level contribution
from contraction with the external state would be boosted by factors of $N$ relative
to quantum loops at each order of perturbation in the boson-fermion interaction.
This gives support to our semi-classical treatment.

Since in the flat space, (charged) fermions induce gauge and
gravitational Chern-Simons terms through parity
anomaly\footnote{Fermion mass term, in fact, breaks parity symmetry
  explicitly in 2+1 dimensions.  On the other hand in 3+1 dimensions
  it does not break parity.}, the result obtained in the
Einstein-Maxwell-Chern-Simons was a precursor to our expectation of
back-reaction of fermions on AdS space.  Armed with this result we
studied fermions in the Schr\"odinger background in Einstein-Maxwell
theory without Chern-Simons terms, anticipating that the Schr\"odinger
spacetime is the fully back-reacted geometry of the fermions in AdS
space.  We indeed found a consistent solution to the
gravity-Maxwell-fermion system in the Schr\"odinger spacetime in
arbitrary dimension for the critical exponent $z= \pm 2mL-(D-2)$, where
$D$ is the number of bulk spacetime dimensions.

The emergence of Schr\"odinger spacetime can also be seen
perturbatively.  Let us start with AdS$_3$ spacetime with metric
\begin{equation}
\label{ads3d}
ds^2=L^2(-4 r dt dx+\frac{dr^2}{4r^2}),
\end{equation}
and consider gauge field components $A_\mu$  taking at most
constant values.  We can then solve the Dirac equation for a fermion
with charge $q$ and mass $m$ in this background.  The solution is
\begin{equation}
  \label{eq:17}
  \psi(r,t,x) = r^{\frac{m}{2} - \frac{3}{4}} \exp(-i\omega t)
  \left(\begin{array}{c}
    1 \\ 1
  \end{array}\right) .
\end{equation}
We then look at the back-reaction of this solution on the background
metric and gauge fields.  Since we are treating this fermion solution
perturbatively, the back-reaction will be tiny (proportional to
$\hbar$) but indicative of the effect of macroscopic droplet of these
fermion on the background.  With this in mind we evaluate fermion
current and energy momentum tensor and feed it back into equations of
motion of the background.  Solution to equations with back-reaction of
fermions generate non-trivial gauge field component $A_t$ and metric
component $g_{tt}$, which take the form
\begin{eqnarray}
  \label{eq:18}
  A_t(r) &=& \hbar ( c_2 + c_1\ln r + \frac{4q
    r^{m-1/2}}{(2m-1)^2}) \\
  g_{tt} &=& \hbar \frac{1}{(2m-1)^2}\left(\frac{-4 q^2
      r^{2m-1}}{(2m^2-3m+1)} + e_1 \ln r + e_2 r^{m-1/2} \ln r \right),
\end{eqnarray}
where $c_i$ are constants of integration and the $e_i$'s are fixed in
terms of $c_i$ via Einstein equation.  In particular, setting $c_1=0$
sets both $e_i=0$.  In this limit we are actually reduced to the
original solution we found previously in (\ref{sch3d}), (\ref{At}) and
(\ref{zzsol}). The fermions only appear differently because we
have picked a different set of vierbeins to suit our purpose here
starting off from a off-diagonal metric.

It is interesting to note that for $z>0$ the Schr\"odinger backgrounds are
horizon-free and as a result do not have macroscopic entropy
associated with them.  The phenomenon of disappearance of the horizon
due to fermion back-reaction is similar to what was proposed in
\cite{hpst}.
It would be interesting to find a dual description of our solutions.
In the 2+1 dimensions we studied the back-reacted geometry supported
either by the Chern-Simons terms or by fermions.  As mentioned earlier,
around flat backgrounds it is known that Chern-Simons terms are
induced by one loop effects in the fermionic theory\cite{red,bb}.  In the context
of curved space we have two Chern-Simons terms, gauge and
gravitational.  The Schr\"odinger metric obtained in the
Chern-Simons theory has (up to sign) two new exponents apart from the
standard one corresponding to AdS space; $z= \mp
4\mu_F$ and $z= (\mu_G\mp 1)/(2\mu_G)$ (see (\ref{csthree})).  On the other
hand, in the fermionic theory we have only one exponent (up to a sign of
 the fermion mass term).  It is tempting to conjecture that even in
curved spacetimes, the fermions would induce Chern-Simons terms.
It would be interesting to directly relate these two theories and
compare their resulting exponent $z$.

Let us make a brief comment on the Maxwell-Chern-Simons theory. Around
flat spacetime it is possible to arrange the coefficient of gauge
Chern-Simons term and the number of fermion species to switch off the
gravitational Chern-Simons term\cite{bpr}. It turns out that this
system has already been studied in \cite{clem,clement} and
\cite{bhnogo}.  In \cite{bhnogo}, it was proved that there exist no
black hole solutions in this theory.  The existence of Chern-Simons
term in 2+1 dimensions implies the presence of a chiral anomaly in the 1+1
dimensional theory \cite{niemi}.  The induced chiral anomaly in the
boundary theory is canceled by the chiral current in the boundary
theory.  In \cite{clement}, it was found (after transforming to our
conventions), that regular black hole solutions only exist if $\mu_F
\mu_G > 1/6$.  This is reminiscent of the condition (\ref{machrg}).

It has been shown in \cite{hpst} that Lifshitz metrics are solutions
when a WKB approximation is made. We leave it as an open problem to
examine if they are also exact solutions. In light of our analysis
using the Chern-Simons theory and the self-gravitating fermions, this
is unlikely to happen in 2+1 dimensions. However in higher dimensions,
there are more ways to solve the constraint $j_r=0$ than in 2+1
dimensions. So it could very well be that both Lifshitz and
Schr\"odinger metrics exist as exact solutions. It will be exciting to
find new backgrounds from self-gravitating fermions. Of course in
general, this is a very difficult problem.

\section*{Acknowledgments}{We would like to thank Sumit Das, Suresh
  Govindarajan, T. Padmanabhan, Sumathi Rao, Harvey Reall, Ashoke Sen,
  Yanwen Shang, Brian Shieh and Subir Sachdev for discussions.
  We also thank Sean Hartnoll and
  David Tong for numerous email exchanges which helped us understand
  what we are doing better.  We would particularly like to thank Rob
  Myers for detailed discussions and for several excellent suggestions.
  We have made some use of the free Mathematica package,
  tools of tensor calculus, in our computations.
  Research at Perimeter Institute is supported by the Government of
  Canada through Industry Canada and by the Province of Ontario
  through the Ministry of Research \& Innovation.}

\begin{appendix}
\setcounter{equation}{0}
\section{Conventions}
\label{sec:conventions}
We use the following conventions for three and four dimensions.  The
three dimensional action with Chern-Simon terms is
\begin{equation}
  \label{eq:19}
  I_3 = \frac{1}{2\ell_p}\int d^3x\left[ \sqrt{-g} (R + \frac{2}{L^2}
    - \frac{1}{4}F_{\mu\nu}F^{\mu\nu}) + L_{MCS} + L_{GCS}\right]
\end{equation}
where the Maxwell Chern Simons and gravitational Chern Simons are given by
\begin{eqnarray}
  L_{MCS}&=&\mu_F \epsilon^{\alpha \beta\gamma}A_\alpha F_{\beta\gamma}\,,\\
  L_{GCS}&=&\mu_G \epsilon^{\lambda \mu \nu}\Gamma^\alpha_{\lambda\beta}
  ( \partial_\mu \Gamma^\beta_{\alpha\nu}+ \frac{2}{3}
  \Gamma^\beta_{\mu\gamma} \Gamma^\gamma_{\nu\alpha})\,.
  \\
\end{eqnarray}
The four dimensional action is
\begin{equation}
  \label{eq:20}
  I_4 =  \frac{1}{2\ell_p^2}\int d^4x \left[ \sqrt{-g} (R +
    \frac{6}{L^2} - \frac{1}{4}F_{\mu\nu}F^{\mu\nu})\right].
\end{equation}
The Einstein equation is
\begin{equation}
  \label{eq:29}
  R_{\mu\nu} - \frac{1}{2}g_{\mu\nu} R +\Lambda g_{\mu\nu} =  \frac{1}{2}T_{\mu\nu},
\end{equation}
where $\Lambda$ is the cosmological constant in appropriate dimensions
and $T_{\mu\nu}$ is the matter stress energy tensor.  The matter
stress energy tensor with this choice of conventions is defined as
\begin{equation}
  \label{eq:30}
  T_{\mu\nu} = -\frac{2}{\sqrt{-g}} \frac{\delta S}{\delta g^{\mu\nu}}.
\end{equation}
The generic asymptotically AdS metric is written as
\begin{equation}
  \label{eq:21}
  ds^2 = - \frac{r^2}{L^2}f(r) dt^2 + \frac{L^2}{r^2}\frac{dr^2}{f(r)}
  + \frac{r^2}{L^2} dx_i^2\, .
\end{equation}
In absence of Chern-Simons terms in three dimensions it can be shown
that (\ref{eq:21}) solves equations of motion to give charged BTZ
black hole solution\cite{Banados:1992wn,Clement:1993kc,Martinez:1999qi}  with
\begin{equation}
  \label{eq:22}
  f(r) = 1- \frac{q^2L^2}{4r^2} -
  \frac{q^2L^2}{2r^2}\ln \frac{r}{r_*},\quad A_t = q \ln \frac{r}{r_0},
\end{equation}
where $r_0$ is the location of outer horizon.  The condition for
existence of two horizons is $q^2L^2 \geq 4r_*^2$ and the
extremality condition is $ r_* = qL/2$.

In four dimensions, (\ref{eq:21}) solves the equations of motion to
give charged Reissner-Nordstr\"om black hole solution, with behaviour
of $f(r)$ and $A_t(r)$ given by
\begin{equation}
  \label{eq:26}
  f(r) = 1 -\frac{M}{r^3} + \frac{q^2L^2}{4r^4},\quad A_t(r) = q (
  \frac{1}{r} - \frac{1}{r_0}).
\end{equation}
In order to have double horizon $27M^4 \geq 4q^6L^6$ and the equality
corresponds to extremal solution.

The near horizon metric in the extremal limit is given by AdS$_2\times
M_n$, where $M_n$ is an $n$-dimensional manifold, which in our case is
either $R^n$ or $S^n$.  Our convention for AdS$_2$ metric in Poincare
coordinates  is
\begin{equation}
  \label{eq:27}
  ds^2_{AdS2} = - \frac{r^2}{\tilde L^2}dt^2 + \frac{\tilde L^2}{r^2} dr^2,
\end{equation}
where $\tilde L$ is the radius of AdS$_2$. In 2+1 dimensions,
according to our conventions,
\be
A_t=\sqrt{2}\frac{r}{\tilde L}\,,\quad \tilde L=\frac{L}{\sqrt{2}}\,,
\ee
while in 3+1 dimensions,
\be
A_t=\sqrt{2}\frac{r}{\tilde L}\,,\quad \tilde L=\frac{L}{\sqrt{6}}\,.
\ee

\setcounter{equation}{0}
\section{Spin connection and the Dirac Equation}
Consider a general metric of the form
\begin{equation}\label{schm}
ds^2=-g_{tt}(r) dt^2+g_{rr}(r) dr^2+ g_{xx}(r) dx^2+ g_{ii}(r) dx^i dx^i + 2g_{tx}(r) dt dx\,,
\end{equation}
with all metric components having only radial dependence.  We will use
$\mu, \nu, \cdots$ to denote curved spacetime indices and $i,j,\cdots$
to denote curved spatial indices.  We will reserve index `$d$' for
denoting flat coordinate related to curved index $r$ by vielbein.
Flat space indices for boundary coordinates are denoted by
$a,b,\cdots$.

In this case, following choice of vielbeins is suitable for both
diagonal metrics as well as Schr\" odinger metrics
\begin{eqnarray}
  \label{eq:1}
  && e^0 = \sqrt{g_{tt}} dt - \frac{g_{tx}}{\sqrt{g_{tt}}}dx,\, \,  e^d
  = \sqrt{g_{rr}} dr \nonumber \\
  && e^1 = \frac{\sqrt{g_{tt}g_{xx}+g_{tx}^2}}{\sqrt{g_{tt}}}dx,\quad
  e^a = \sqrt{g_{ii}} dx^i,\, a=2,3,\cdots .
\end{eqnarray}
It is straightforward to derive components of spin connections from
them and they are given by
\begin{eqnarray}
  \label{eq:2}
  &&\omega_{d0} = \frac{1}{2} \frac{g'_{tt}(r)}{\sqrt{g_{tt}g_{rr}}}dt
  -\frac{1}{2}\frac{g_{tx}}{\sqrt{g_{tt}g_{rr}}}
  \frac{g'_{tt}(r)}{g_{tt}}dx, \\
  && \omega_{d1} = - \frac{1}{2\sqrt{(-\det g)g_{tt}}}\left[ ( g_{tt}
    g'_{xx}(r) + g_{tx} g'_{tx}(r)) dx + ( g_{tt} g'_{tx}(r) - g_{tx}
    g'_{tt}(r))dt \right],\\
  && \omega_{di} = \frac{g_{ii}'(r)}{2\sqrt{g_{rr}g_{ii}}}dx^i,
\end{eqnarray}
where primes one the metric components are derivatives with respect to
their argument.

Let us now look at the Dirac equation in different backgrounds.  The
Dirac equation for a particle with mass $m$ and charge $e$ takes the
form
\begin{equation}
  \label{eq:24}
  (/\!\!\!\!D - m)\psi = 0,
\end{equation}
where
\begin{equation}
  \label{eq:25}
  /\!\!\!\!D = \Gamma^c e_c^\mu(\partial_\mu +
  \frac{1}{4}\omega_{\mu ab} \Gamma^a\Gamma^b + i q A_\mu).
\end{equation}
To solve the Dirac equation we will use the following $\Gamma$-matrix
convention for 2+1 dimensions,
\begin{equation}
  \label{eq:9}
  \Gamma^1= \sigma_3,\quad \Gamma^0 = i \sigma_2,\quad \Gamma^d =
  \sigma_1\, .
\end{equation}
Our 3+1 dimensional $\Gamma$-matrix convention is
\begin{equation}
  \label{eq:23}
  \Gamma^d = \left(
    \begin{array}{cc}
      I & 0\\ 0 & -I
    \end{array}
\right),\quad \Gamma^a = \left(
  \begin{array}{cc}
    0 & \sigma^a \\ \sigma^a & 0
  \end{array}
\right),\quad a = 0, 1, 2
\end{equation}
Where $I$ is $2\times 2$ identity matrix and $\sigma^a$ are
Pauli matrices.

\section{Stress tensor of fermionic probes in AdS}
\label{sec:stress-tens-ferm}
\setcounter{equation}{0}

We will be considering fermions in curved space. We shall begin
presenting{\footnote{Note that \cite{birrell} use the mostly negative
    convention while we follow the more standard GR convention of
    mostly positive.}} explicitly the fermionic contribution to the
stress tensor \cite{birrell}.
The quadratic action for a charged massive Dirac spinor is
\begin{equation}
  S_{\rm Dirac}= \frac{1}{2\ell_p^{d-1}} \int d^{d+1}x\sqrt{-g}i(\bar\psi\Gamma^ae_a^\mu
  D_\mu\psi-m\bar\psi\psi)
\end{equation}
where
\begin{eqnarray}
&&D_\mu \psi=(\partial_\mu+
\frac{i}2\omega_\mu^{ab}\Gamma_{ab}+iqA_\mu)\psi,
\qquad  \bar{\psi} \overleftarrow{D}_{\mu}=(\partial_\mu \bar{\psi}-
\frac{i}2\omega_\mu^{ab}\bar{\psi}\Gamma_{ab}
-iqA_\mu \bar{\psi}), \nonumber \\
&&\bar\psi=\psi^\dagger \Gamma^0, \qquad \Gamma_{ab}= -
\frac{i}4[\Gamma_a,\Gamma_b]
\end{eqnarray}
The corresponding stress tensor is
\be \label{fermT}
T^f_{\mu\nu} = \frac{1}{2}\left( - i \bar{\psi} \Gamma_{(\mu}D_{\nu)} \psi
+ i  \bar{\psi}\overleftarrow{D}_{(\mu}  \Gamma_{\nu)}\psi \right),
\ee
and symmetrization is denoted by
$V^1_{(\mu} V^2_{\nu)}= \frac{1}{2} (V^1_\mu V^2_\nu + V^1_\nu V^2_\mu)$.

\subsection{AdS$_2\times R$}
\label{sec:ads_2times-r}

The background is, following our conventions,
\be
ds^2_{AdS2\times R} = - \frac{r^2}{\tilde L^2}dt^2 +
\frac{\tilde L^2}{r^2} dr^2 + dx^2, \qquad A_t = \frac{\sqrt{2} r}{\tilde{L}},
\ee
and the only non-vanishing spin connection is,
\be
 \omega_{10} = \frac{r}{\tilde L^2} dt .
\ee
Taking a Dirac spinor of the form
\be \label{3dspinor}
\psi = \left(\begin{array}{c}\psi_+\\\psi_-\end{array}\right),
\ee
and assuming only $r$ dependence, the Dirac equation is given explicitly by
\be
2r \psi_\pm' + 2i q\sqrt{2}\psi_\pm + (1\mp 2m\tilde{L})\psi_\mp =0.
\ee
The solution is simply
\be
\psi^{\pm} = r^{-\frac{1}{2}\pm \delta}\left(
\begin{array}{c} p^{\pm}_1\\p^{\pm}_2 \end{array}
\right),\qquad p^{\pm}_2 = \pm   \frac{i p^{\pm}_1
(\delta \mp m\tilde{L} )}{\sqrt{2}L q}, \qquad \delta =
\tilde{L} \sqrt{m^2 - 2q^2},
\ee
for some constants $p^{\pm}_+$.
Switching on one of the two solutions leads to
\begin{eqnarray}
T^f_{tt} &=&  \frac{|p^{\pm}_1|^2 \sqrt{2} m (m \tilde{L}
\mp \delta)}{ \tilde{L}^3 q} r^{1\pm 2\delta},\nonumber \\
T^f_{xt} &=& \pm |p^{\pm}_1|^2\frac{m\tilde{L}(2\delta
\mp 1) +\delta \mp 2\delta^2}{2\sqrt{2}q \tilde{L}^3}.
\end{eqnarray}
Note the appearance of $T_{xt}$ components. The $rr$ component
contributes only from cross terms when both solutions are switched
on.  Note that the Ricci tensor for the background metric is simply
\be
R_{tt} = \frac{r^2}{\tilde{L^4}},\qquad R_{rr} = -\frac{1}{r^2}.
\ee
Therefore, in the near horizon limit $r\to 0$, the solution scaling
with $r^{-(\frac{1}{2}+ \delta)}$ would eventually become dominant
over the background such that the probe approximation breaks down.
For the other solution that scales as $r^{-(\frac{1}{2}- \delta)}$
however, the probe approximation is good all the way to the horizon if
$\delta >1$, which gives
\be
\label{machrg}
m^2 >2q^2+ 1.
\ee
When the mass is smaller than $2q^2 +1$, back reaction in the near
horizon region cannot be ignored, with the likely result of destroying
the horizon altogether. Therefore in the large mass and charge limit
we recover the assertion in \cite{hpst} that roughly horizon is
destroyed whenever $eq>m$, where $e=\sqrt{2}$ is the background
electric field.
Note also that these solutions independently satisfy
\be
T^t_t - T^r_r = - \frac{|p^{\pm}_1|^2 \sqrt{2} m (m \tilde{L}
\mp \delta)}{ \tilde{L} q} r^{-1\pm 2\delta} <0.
\ee

\subsection{AdS$_2\times R^2$}
  \label{sec:ads_2times-r2}

Let us repeat the exercise in the previous section in AdS$_2 \times R^2$.
A static solution of a 4-component Dirac fermion would be given by
\begin{equation}
\psi = \psi_+ r^{-\frac{1}{2}+ \delta} + \psi_-
r^{-\frac{1}{2}-\delta},
\qquad \delta = \tilde{L} \sqrt{m^2 - 2q^2},
\end{equation}
where
\begin{equation}
\psi_\pm = \left(\begin{array}{c}\psi^1_\pm\\ \psi^2_\pm
\end{array}\right),\qquad \psi^2_\pm = \pm
\frac{(\frac{\delta}{\tilde{L}}
\mp m  )}{\sqrt{2}q \tilde{L}}(\sigma_1 \psi^1_\pm),   \label{fermsol1}
\end{equation}
where $\psi^i_\pm$ are two component constant spinors and $\sigma_i$
are the Pauli matrices.
The non-vanishing contribution of $\psi_+$  to the energy-momentum tensor is given by
\begin{eqnarray}
T^f_{tt} &=& \psi^{1\dagger}_\pm\psi^1_\pm  \frac{ r^{1 \pm 2 \delta}
(\tilde{L}^2 (m^2 + 2 q^2) \mp 2 \tilde{L} m \delta +
\delta^2)}{\sqrt{2}\tilde{L}^4 q} \\
T^f_{tx} &=& \psi^{1\dagger}_\pm \sigma_3 \psi^1_\pm
\frac{ r^{\pm 2 \delta} (\tilde{L}^2 (m^2 - 2 q^2) + (\delta\mp1) \delta +
   \tilde{L} (m \mp 2 m \delta))}{2 \sqrt{2} \tilde{L}^3 q}  \\
T^f_{ty} &=& -\psi^{1\dagger}_\pm \sigma_2 \psi^1_\pm
\frac{r^{\pm 2 \delta} (\tilde{L}^2 (m^2 - 2 q^2) + (\delta\mp1) \delta +
   \tilde{L} (m \mp 2 m \delta))}{2 \sqrt{2} \tilde{L}^3 q}\\
\end{eqnarray}
As in the previous section, the energy-momentum clearly illustrates
again that back-reaction is severe in the near horizon region
generally for $\sqrt{2}q>m $
Note that here
\be\label{nullads2r2}
T^t_{\ t} - T^r_{\ r} = -\psi^{1\dagger}_\pm\psi^1_\pm
\frac{ r^{-1 \pm 2 \delta} (\tilde{L}^2 (m^2 + 2 q^2) \mp 2
\tilde{L} m \delta + \delta^2)}{\sqrt{2}\tilde{L}^2 q} <0
\ee

\subsection{AdS$_2\times S^2$}
\label{sec:ads_2times-s2}

An entirely parallel story goes through if one were to replace $R^2$
in the previous sub-section by $S^2$.  Namely
\begin{equation}
\psi = \psi_+ r^{-\frac{1}{2}+ \delta} + \psi_-
r^{-\frac{1}{2}-\delta},
\qquad \delta = \tilde{L} \sqrt{m^2 - 2q^2},
\end{equation}
where, however even the lowest mode contains non-trivial angular dependence,
\begin{equation}
\psi_\pm = \frac{1}{\sqrt{\sin \theta}}\left(\begin{array}{c}
\psi^1_\pm\\ \psi^2_\pm
\end{array}\right),\qquad \psi^2_\pm = \pm
\frac{(\frac{\delta}{\tilde{L}} \mp m  )}{\sqrt{2}q
\tilde{L}}(\sigma_1 \psi^1_\pm),   \label{fermsol2}
\end{equation}
The energy-momentum tensor here is similarly,
\begin{eqnarray}
T^f_{tt} &=& \csc \theta \psi^{1\dagger}_\pm\psi^1_\pm
\frac{ r^{1 \pm 2 \delta} (\tilde{L}^2 (m^2 + 2 q^2) \mp 2 \tilde{L}
m \delta + \delta^2)}{\sqrt{2}\tilde{L}^4 q} \\
T^f_{t\theta} &=& \csc \theta \psi^{1\dagger}_\pm \sigma_3 \psi^1_\pm
\frac{ r^{\pm 2 \delta} (\tilde{L}^2 (m^2 - 2 q^2) + (\delta\mp1) \delta +
   \tilde{L} (m \mp 2 m \delta))}{2 \sqrt{2} \tilde{L}^3 q}  \\
T^f_{t\phi} &=& \frac{-\sqrt{2}\tilde{L}\cos\theta}{8q r}T^f_{tt} -
\psi^{1\dagger}_\pm \sigma_2 \psi^1_\pm
\frac{r^{\pm 2 \delta} (\tilde{L}^2 (m^2 - 2 q^2) + (\delta\mp1) \delta +
   \tilde{L} (m \mp 2 m \delta))}{2 \sqrt{2} \tilde{L}^3 q}\\
\end{eqnarray}
The null energy condition is the same as eq. (\ref{nullads2r2}).

\section{A toy model justification for our
semi-classical treatment}
\setcounter{equation}{0}

Our discussion is largely based on the assumption that in the presence
of a large fermi-gas, one can treat the background classically, while
the fermions propagate in this mean-field background
quantum-mechanically.  We consider here a toy model where fermions are
coupled to bosons. In the case where the fermion fields appear only up
to quadratic order, we can obtain the 1-loop effective action of the
bosons after integrating out the fermions in an $N$-fermion state. It
turns out, in the large $N$-limit and to leading order in the
boson-fermion coupling $g$, the effective equations of motion of the
bosons are equivalent to the original one, except the fermion
operators are replaced by their expectation values evaluated on the
$N$-fermion state. The toy model gives us more confidence in the
methods we advocate in this paper. It is also noteworthy that the
approach of treating back-reacting fermions on a mean-field background
and obtaining self-consistent set of solutions to all the equations of
motion with an explicit built-in Fermi-surface, is a well-known
strategy in nuclear physics\cite{Gambhir:1989mp}, which is equivalent
in various limits to the Hartree-Fock approximation.

Our discussion here will follow closely \cite{kapusta}. The
(Euclidean) action of our toy-model at temperature $T=\beta^{-1}$in
(0+1) dimension is as follows:
\begin{eqnarray}
&&S = S_b + S_f + S_{int}, \qquad S_b =\frac{1}{2} \int_0^\beta
d\tau \dot{\phi}^2 + m_b^2 \phi^2, \nonumber \\
&&S_f + S_{int} =
\int_0^\beta d\tau \sum_i  b_i^\dag \dot{b}_i +(m_{i}-
\mu_i )b^\dag_ib_i - g \phi b_i^\dag b_i,
\end{eqnarray}
where we have introduced many species of fermionic fields $b_i$ and a
corresponding chemical potential $\mu_i$ for each species, which are
to be determined from some given occupation of the fermionic states.
The partition function is given by
\be
Z = \int D[\phi] \prod_i D[i b^\dag_i]D[b_i] \exp(\frac{- S}{\hbar}).
\ee
Since the action is quadratic in the fermions, the path-integral can
be readily done. Expressing the fields as
\be
b_i = \sum_n e^{-i\omega^f_n \tau} b_{i;n},\qquad
\phi= \sum_n e^{-i\omega^b_n \tau} \phi_n
\ee
The $\omega_n$ here are thus Matsubara frequencies given by
\be
\omega^f_n = (2n+1)\pi T, \qquad \omega^b_n = 2n\pi T,
\ee
and
$S_f + S_{int}$ can be re-written as
\be
S_f+ S_{int} = \sum_{i,m,n} i b^\dag_{i;m}\mathcal{D}^i_{mn}
b_{i;n},\qquad \mathcal{D}^i_{mn}= -i\beta((-i\omega_n +
\mu_i - m_i)\delta_{mn} - g \phi_{m-n} ).
\ee

Integrating out the fermions one is then left with
\be
Z = \int D[\phi] \exp(-\frac{S_b}{\hbar}) \det(\frac{\mathcal{D}}{\hbar}) = \det(\beta(-i\omega_n + \mu_i - m_i)) \int D[\phi] \exp(-\frac{S_b}{\hbar}) \det(1- \frac{g\phi_{m-n}}{-i\omega_n + \mu_i - m_i}).
\ee
To lowest order in the coupling $g$, the fermion contribution to the free energy evaluates to \cite{kapusta}, using various summation formula,
\begin{eqnarray}
\frac{F_f}{T} &&= - \ln Z_f = -\sum_{n,i} \ln [\beta (-i\omega_n + \mu_i - m_i)] \nonumber \\
&&= -\frac{1}{2}\sum_i [\beta^2 (\sum_n (\omega^2_n + (\mu_i- m_i)^2))] \nonumber \\
&&= -\sum_i  \ln (1+ e^{-\beta(m_i-\mu)}),
\end{eqnarray}
where we have dropped the contribution of vacuum energy which has to be regulated.
We can determine the chemical potentials by
\be
\partial_{\mu_i}(T \ln Z) \sim \frac{1}{e^{\beta(m_i - \mu_i)}+1} = N_i.
\ee
which is precisely the Fermi-distribution. The ground state
\be|Q\rangle = \prod_i b^\dag_i |0\rangle,
\ee
corresponding to a system of $N$ fermions with fermi-energy $E_F$ can thus be obtained
by picking
\be
\mu_i = E_F.
\ee
In which case the Fermi-distribution behaves like a step function, giving occupation $N_i=1$ for all single particle
states below $E_f$. Now returning to the full path integral, we have
\begin{eqnarray}
Z &&= \int D[\phi]\exp(-\frac{S_b}{\hbar}) \exp(\ln \det(\frac{D}{\hbar}))  \nonumber \\
&&\sim \int D[\phi]\exp(-\frac{S_b}{\hbar})\exp(\sum_i \beta m_i + \ln (1+ e^{-\beta(m_i-E_f - g\phi_0)})- \ln \hbar),
\end{eqnarray}
where in the second line we have taken the leading $g$ approximation,
so that in the determinant of a matrix of the form
\be
\left(\begin{array}{ccc}(1+ g\phi_{0})& g\phi_{1}&\ldots\\
                         g\phi_{-1}  & (1+ g\phi_{0})&\ldots\\
                         \vdots & \vdots& \ddots \end{array}\right),
\nonumber
\ee
the leading $g$ dependence came from the trace, which include only the
diagonal components with $\phi_{(m-m=0)}$. The off-diagonal components
can only arise in higher $g$ corrections.

Since in the zero-temperature limit $\beta\to \infty$ the factor
$e^{-\beta(m_i-E_f)}$  approaches infinity for $m_i <E_f$ and zero
otherwise, we have
\begin{eqnarray}
Z &&\sim \int D[\phi]\exp(-\frac{S_b}{\hbar})\exp(
\sum_{i=\textrm{filled states}} -\beta(m_i-E_f - g\phi_0)) \nonumber \\
&&\sim \int D[\phi]\exp(-\frac{S_b}{\hbar})\exp( -\beta(N g\phi_0) + \ldots),
\end{eqnarray}
where the vacuum energy that is supposed to be regulated, and other
pieces independent of $\phi$ are denoted in $\ldots$. Note that to
leading order in $g$ the effective action of the bosons after
integrating out the fermions, is precisely as if we replace the
fermion operators by their expectation values in the $N$-fermion
state. For back-reaction on the bosons to be important however, we
need
\be\label{Ng}
N g \sim \frac{1}{\hbar}.
\ee
Note also that the approximation we are taking in the partition
function is equivalent to, from the perspective of perturbation theory
via Feynman diagrams, only the tree diagrams i.e. The perturbation
operators are only contracting with external states. Consider for
example the leading order $g$ correction to any expectation values of
some arbitrary operators $\mathcal{O}$ in the presence of the
interaction terms,
\be
\langle Q|\mathcal{O} \int \phi \sum_i b_i^\dag b_i |Q\rangle,
\ee
we see that contraction of the $b_i$'s with the external creation
operators in $|Q\rangle$ result in a factor of $N$, whereas
contraction among the $b_i$'s, corresponding to loop diagrams, is only
of order 1 relatively. Similarly at order $g^2$, we have
\be
\frac{1}{2!}\int\int  \sum_{i,j} \phi \phi \langle Q|
\mathcal{O} (b_i^\dag b_i)  ( b_j^\dag b_j) |Q\rangle,
\ee
where contraction of the $b$'s with the external state would again
give rise to an extra factor of roughly $N(N-1)\sim N^2$ relative to
contraction among the $b$ and $b^\dag$.\footnote{ Note that the
  combinatoric factor of $2!$ would appear in either internal
  contractions or contraction with the external state. We have also
  made used of the fact that states with different excitations are
  orthogonal to each other. Therefore having chosen a particular
  contraction for the $b_j$'s on $|Q\rangle$, the contraction of
  $b_i^\dag$ with $\langle Q|$ is completely determined. This explains
  why there is only a factor of $N(N-1)$, rather than the square of
  it.} In the limit where $N$ is large and $g$ is small, the loop
diagrams would be suppressed relative to tree diagrams.

To summarize, in the limit where $Ng\sim \hbar^{-1}$, the
back-reaction of the fermi-gas on the bosons become significant, but
at the same time we have the quantum loops under control because $g$
is small. Particularly, the leading $g$ effect on the bosons can be
captured by replacing all the fermionic operators by their expectation
values evaluated on the filled state $|Q\rangle$ in the equations of
motion of the bosons. In fact the coupling to the bosonic field $g
\phi b^\dag b$ could have been generalized to some arbitrary one of
the form $gV(\phi)b^\dag b$ without altering the conclusion, as long
as $g$ is small. The toy model appears to justify the approach taken
in this paper in including fermionic back-reactions on a bosonic
background.

\end{appendix}

\end{document}